\documentclass[sigconf,screen]{acmart}

\AtBeginDocument{%
  \providecommand\BibTeX{{%
    \normalfont B\kern-0.5em{\scshape i\kern-0.25em b}\kern-0.8em\TeX}}}

\usepackage{fontawesome}
\usepackage{framed}

\usepackage{xspace}
\usepackage{soul}
\usepackage{upgreek}
\usepackage{euscript}
\usepackage{booktabs}
\usepackage{siunitx}

\usepackage{adjustbox}
\usepackage{flushend}
\usepackage{balance}
\usepackage{multirow}
\usepackage{subcaption}

\usepackage[normalem]{ulem}

\usepackage{ragged2e}

\usepackage{booktabs}
\usepackage{array}
\newcolumntype{P}[1]{>{\raggedright\arraybackslash}p{#1}}

\begin{document}

\title[\faicon{volume-up} Everything We Hear: Towards Tackling Misinformation in Podcasts]{\faicon{volume-up} Everything We Hear: \\Towards Tackling Misinformation in Podcasts}


\author{Sachin Pathiyan Cherumanal}
\orcid{https://orcid.org/0000-0001-9982-3944} 
\affiliation{
\institution{RMIT University}
\city{Melbourne}
\country{Australia}
}
\email{s3874326@student.rmit.edu.au}

\author{Ujwal Gadiraju}
\orcid{https://orcid.org/0000-0002-6189-6539}
\affiliation{
\institution{Delft University of Technology}
\city{Delft}
\country{The Netherlands}
}
\email{u.k.gadiraju@tudelft.nl}

\author{Damiano Spina}
\orcid{https://orcid.org/0000-0001-9913-433X}
\affiliation{
\institution{RMIT University}
\city{Melbourne}
\country{Australia}
}
\email{damiano.spina@rmit.edu.au}

\copyrightyear{2024}
\acmYear{2024}
\setcopyright{rightsretained}
\acmConference[ICMI '24]{International Conference On Multimodal Interaction}{November 4--8, 2024}{San Jos\'e, Costa Rica}
\acmBooktitle{International Conference On Multimodal Interaction (ICMI '24), November 4--8, 2024, San Jos\'e, Costa Rica}\acmDOI{10.1145/3678957.3678959}
\acmISBN{979-8-4007-0462-8/24/11}
\makeatletter
\gdef\@copyrightpermission{
  \begin{minipage}{0.3\columnwidth}
   \href{https://creativecommons.org/licenses/by-nd/4.0/}{\includegraphics[width=0.90\textwidth]{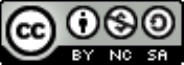}}
   \end{minipage}\hfill
   \begin{minipage}{0.7\columnwidth}
\href{https://creativecommons.org/licenses/by-nd/4.0/}{This work is licensed under a Creative Commons Attribution-NoDerivs International 4.0 License.}
  \end{minipage}
  \vspace{5pt}
}
\makeatother
\begin{abstract}

Advances in generative AI, the proliferation of large multimodal models (LMMs), and democratized open access to these technologies have direct implications for the production and diffusion of misinformation. In this prequel, we address tackling misinformation in the unique and increasingly popular context of podcasts. The rise of podcasts as a popular medium for disseminating information across diverse topics necessitates a proactive strategy to combat the spread of misinformation. Inspired by the proven effectiveness of \textit{auditory alerts} in contexts like collision alerts for drivers and error pings in mobile phones, our work envisions the application of auditory alerts as an effective tool to tackle misinformation in podcasts.
We propose the integration of suitable auditory alerts to notify listeners of potential misinformation within the podcasts they are listening to, in real-time and without hampering listening experiences. We identify several opportunities and challenges in this path and aim to provoke novel conversations around instruments, methods, and measures to tackle misinformation in podcasts.

\end{abstract}


\maketitle

\section{Introduction}
\label{sec:introduction}

Podcasts have shown to have a growing listenership over the recent years \cite{jones2021podcastschallenges, tian2022spotify}. As of 2024, there are 464.7 million global podcast listeners. This number is predicted to reach 504.9 million by 2024~\cite{demandsage}.
In today's world, podcasts have become a major source of information on a variety of topics such as politics, current affairs, culture, health, and controversial issues. Surveys have shown that people listen to podcasts while completing mundane tasks such as commuting, and household chores \cite{tobin2022people}. It is important to note that podcasts may sometimes unintentionally spread misinformation. One such example is when Joe Rogan's influential podcast hosted infamous physician Dr. Robert Malone promoting COVID-19 conspiracy theories (Refer \cite{caramancion2022exploration}).  As a result, such podcasts may end up amplifying inaccurate or biased information to their listeners, which can negatively impact society. The democratization of podcasts and the lack of editorial oversight, fact-checking, or accountability mechanisms in the podcasting ecosystem may contribute to the spread of misinformation. Furthermore, podcast users are loyal listeners and this only exacerbates the issue of misinformation unless users are made aware \cite{doring2022listen}.  This makes the problem of tackling misinformation still an open challenge in audio-based information access systems such as spoken conversational search (where communication between a user and system occurs verbally through audio) and podcasts.

Fact-checking is a long-standing practice in journalism that publishes an evidence-based analysis (i.e., reports) of the accuracy of a political claim, news report, or other public text~\cite{graves2019fact}. Previous studies \cite{porter2021global} show it's effective against online misinformation. However, fact-check reports can be complex and require detailed reading on a screen. Recent research has explored different presentation strategies~\cite{hettiachchi2023factcheckingreportpresentation,konstantinou2019combating}.
Our focus is on exploring techniques to alert users to the existence of fact-checking reports verifying specific claims mentioned in podcasts. In a visual presentation, this can be accomplished with a hyperlink, citation, or even warning labels \cite{martel2023misinformation,pennycook2020implied}. However, in audio-only scenarios like podcasts, where interruptions are not viable (to sustain the train of thought of listeners), finding ways to inform users about potential misinformation and the availability of associated fact-checking reports is crucial. In our work, the goal of these alerts/interventions is to signal potential misinformation and promote critical thinking. 

Auditory interventions have been shown to act as a vital corollary to visual warnings \cite{edworthy1999learning} and cues can be useful especially when the user is engaged in other tasks as is the case with most podcast users \cite{tobin2022people}. Henceforth, in our paper, we discuss the use of auditory intervention techniques to alert podcast listeners about any misinformation contained in certain parts of the podcast. This alert can also be seen as a warning. Such auditory interventions are not a new concept and have been a part of our everyday lives for years. For example, collision warnings in vehicles~\cite{gray2011looming}, mobile notifications, or auditory cues (i.e., interventions) on websites~\cite{batterman2013auditory, murphy2010mathauditory}. However, utilizing auditory interventions is not straightforward.

We propose a novel approach to tackle misinformation in podcasts by harnessing auditory interventions to alert users of potential inaccuracies within snippets of the podcast. We identify the research and empirical gap, discuss open challenges and opportunities, and shed light on potential empirical studies, instruments, and data required to address the current gaps. We hope that this prequel can inspire researchers and practitioners to tackle this important societal problem of misinformation in podcasts.
\section{Background}
\label{sec:related_works}

\subsection{Misinformation, Fact-Checking, and Podcasts}

With the rapid digitization of society, users frequently encounter misinformation, causing harm to life, injury, income, business, emotions, trust, reputation, safety, privacy, decision-making, and more~\cite{li2023combating, tran2020investigation}.
Misinformation has been shown to be multi-modal~\cite{caled2022digital} and exists on various platforms like YouTube \cite{hussein2020measuringmisinformation}, X (Twitter) \cite{ghenai2017healthmisinformation}, search engines \cite{metaxa2017google, pradeep2021misinformationinconsumerhealthsearch, song2022misinformationdensityinsearch}, and podcasts \cite{jones2021podcastschallenges}. Numerous techniques (manual and automated) and tools~\cite{olan2022fake} have been proposed to measure, detect, and mitigate misinformation. However, most work focuses on screen-based interfaces, while addressing misinformation in audio-based formats like podcasts and spoken conversational searches is unexplored~\cite{kiesel2021argumentsearch, ji2024towards}. Fact-checking reports (i.e., manual detection) validated by professionals have been popular method in tackling misinformation, however, such interventions need to be disseminated on the internet wider and faster than fake news to have an impact~\cite{li2023combating}. It is important to note that our work focuses on presenting interventions for misinformation rather than automatically detecting them. In a text-based interface, a fact-checking report can be presented with a hyperlink or citation. However, in audio-only scenarios like podcasts and spoken conversational search, where speech-based interruptions are not viable (to sustain the train of thought of listeners and create good listening experiences), informing users about potential misinformation and availability of associated fact-check reports is crucial but challenging. In this paper, we focus on podcasts, where the interaction is one-way i.e., unlike spoken conversational systems which are interactive. However, the format of a podcast can be quite complex and vary in presentation styles, involving multiple speakers to just a single speaker, i.e., a monologue \cite{carterette2021podcastmetadata}. 
Advancing our empirical understanding would require considering monologue-style podcasts distinctly from multi-speaker settings \cite{spina2015SpeakerLDA}. It is stated that training and warning the users are the two main strategies to improve deception detection, as they lead to improved deception detection accuracy 
\cite{depaulo2003cues, grazioli2004did, ross2018fake}. According to \citet{wogalter2006purposes}, a warning is characterized as a communication medium or a tool that informs about hazards. We can relate this to our context and, misinformation/deceptive content can be thought of as analogous to hazards. There has been empirical evidence indicating that a warning improves individuals’  detection skills \cite{biros2002inducing, Zhang2014effectsofsecuritywarnings}. Furthermore, research has also shown that up-front warnings can reduce later reliance on misinformation \cite{siebert2023effective}. However, these studies focused on visual warnings in websites. While one can argue that visual warnings \cite{kesselheim2015mandatory} help in raising a user's suspicion towards deception, they fail to maintain the user's attention long enough during a podcast. As previous studies have shown attention maintenance is necessary for the receiver to extract the necessary information from the content as well as the warning~\cite{conzola2001communication}. Auditory warnings/interventions can be beneficial (especially in an audio channel like a podcast) because they leverage people’s “always-on” hearing ability. In this work, we highlight the potential for using auditory interventions (e.g., an error ping/a chime on your mobile phone) in podcasts. This raises the following questions: (i) \textit{Can auditory warnings included in podcasts assist users in recognizing misinformation?} (ii) \textit{What kind of auditory warnings are most salient and consistently perceived by different users?} (iii) \textit{What is the general impact of various auditory signals on users, and how does it impact the user's comprehension of the podcast content, interpretation of the signals, and general listening experience?} 

Sections \ref{subsec:related_works_auditory_intervention} and \ref{subsec:related_works_listening_experience} review literature on auditory interventions and experimental methodologies for their impact in podcasts, respectively.

\subsection{Auditory Interventions}
\label{subsec:related_works_auditory_intervention}

\begin{figure}
    \centering
    \includegraphics[trim=2cm 5cm 2cm 5cm,clip,width=0.95\linewidth]{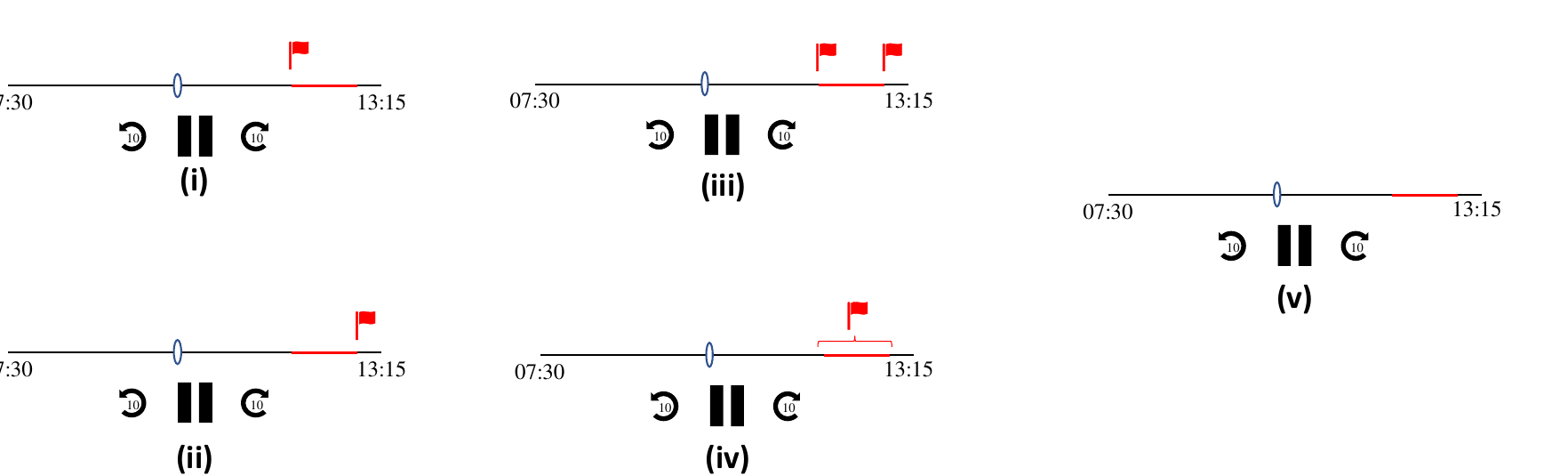}
    \Description[Illustration of five different positioning of auditory icons to identify misinformation in audio podcasts.]{Illustration of five different positioning of auditory icons to identify misinformation in audio podcasts.}
    \caption{Illustrating the different positioning of auditory icons. The red horizontal line represents a section of the podcast that contains misinformation, and the red flag represents the position of the auditory icon. (i) Start of the snippet (S), (ii) End of the snippet (E), (iii) the Start and End of the snippet (S\&E) (iv) Concurrent (C), where the auditory icon would be played concurrently with the snippet that has misinformation and finally (v) No Warning, where no auditory icons will be placed. Note that the red horizontal line and the red flag are for demonstration purposes only, these are not to be displayed to the user. The small blue ellipse denotes the current position on the seek bar.}
    \label{fig:marker_position}
    \vspace{-1em}
\end{figure}

From the starting-up chime of computers to the error pings and mail notifications, auditory interventions have been an integral part of human-computer interaction \cite{Csapo2013overviewauditoryrepresentations, roginska2013auditory}. Such interventions are also popularly used in mission-critical technologies in the automotive industry to warn the driver of a collision \cite{duan2023improving, gray2011looming}, in our mobile phones to notify the users of an incoming email~\cite{garzonis2009notifications}, or as cues to indicate the click of a button~\cite{morrissey2000can}. These auditory warnings can be broadly categorized into speech-based and non-speech-based. It is worth noting that the most recent research on audio interventions (also known as nudges) \cite{Gohsen2023Nudge} explores six different intervention techniques that can be categorized into three groups: (i) linguistic, (ii) natural voice emphasis, and (iii) artificial voice emphasis. It is important to mention that these interventions are primarily speech-based. Although speech is the most semantically rich acoustic medium, it also has a few shortcomings, one of which is privacy concerns, making it the least preferred option for some types of public notifications or reminders~\cite{garzonis2009notifications, baber2002interactive, patterson1990auditory}. On the contrary, non-speech-based warnings have shown more confidentiality, speech independence, and wider applicability in different countries, languages, and dialects \cite{lei2022impact}, making it the most suited for an audio-only context such as podcasts. Non-speech warnings have shown better user task performance ~\cite{ming2008study} and were shown to be preferred over speech-based ones for longer audio content, as the latter can interfere with concurrent speech communication. Recent work by \citet{nees2023auditory} compares such non-speech-based warnings and are classified into three categories\cite{absar2008usability} (i) \textit{auditory icons} -- sounds that can be easily attributed to objects or events generating sounds in everyday situations and (ii) \textit{earcons} \cite{thapa2017nonspeech} -- abstract sounds with no ecological relationship to their referent (target object or event); and (iii) \textit{spearcons} (Speech-Based Earcons) \cite{walker2013spearcons}. Although there have been previous studies comparing auditory icons, earcons, and spearcons, there has been no research on using auditory icons to alert users to misinformation. We propose focusing efforts on auditory icons because previous research suggests they are easier to learn compared to earcons \cite{edworthy2013medical}, and spearcons may interfere with podcast audio content. These auditory icons are further classified into three types based on their relationship with the referent \cite{gaver1987auditory}.\footnote{Here, by referent we mean the target phrase that involves misinformation.} (i) Symbolic: the relation between the sound(s) of the auditory icon and the referent is essentially arbitrary; (ii) Iconic or Nomic: the sound of the auditory icon is related to the physical source of the referent; and (iii) Metaphorical: where the relation between sounds in the auditory icon and the referent is not completely arbitrary, yet also not fully dependent on physical causation. The arbitrariness is based on some similarities between the sound and the referent. \citet{gaver1987auditory} argued that with regard to his classification, in general, iconic/nomic mappings are more powerful than symbolic and metaphorical mappings, because iconic/nomic mappings show a direct relation between the auditory icon and the referent's physical source. This raises the following questions: (i) Which type of auditory icon would be the most effective design choice to implement as a misinformation alert in audio-only content such as podcasts (i.e., iconic/nomic or metaphorical)?\footnote{We exempt the symbolic auditory icon from the study as these can be thought of as reserved non-speech sounds that may have already been conceptually mapped to a specific function of the society (e.g., a police siren).} and (ii) How do we place the auditory icons in a podcast w.r.t. the snippet that contains misinformation? Figure~\ref{fig:marker_position} shows some of the alternatives.

\subsection{Human Factors in Auditory Warnings and Misinformation}
\label{subsec:related_works_listening_experience}

Warnings can be transmitted primarily along two dimensions -- visual (e.g., text, pictures, or both) and audio ( e.g., chimes, pings, voice). A user's response/perception to a warning can vary based on the channel it is transmitted. In this section, we review previous work on human factors in auditory warnings, relevant to our context (i.e., Podcasts). While one of the main advantages of auditory warnings is that it is generally guaranteed that the message will impinge on the receptors in the ear~\cite{conzola2001communication} making sure the users hear it. However, recent studies have also shown auditory feedback may affect a users comprehension~\cite{sheng2020effect}. This warrants a listening comprehension task in future studies that look at auditory warnings in podcasts. This task will involve participants answering questions about the content of the podcast they listened to, as suggested by \citet{pennycook2020fighting}. While we focus on the content of the podcast, it is also imperative we understand, the effect auditory warnings have on users. This raises further questions, such as the following: Can users conceptually map the auditory warnings to misinformation? or do users need to be trained to map these warnings to misinformation? Moreover, a recent work shows auditory warnings may affect working memory~\cite{lei2022impact} , warranting study of their long-term psychological effects.Recent work has also shown that the consumption and interpretation of misinformation can be primarily attributed to cognitive biases like Confirmation Bias\footnote{Confirmation bias is a cognitive bias that causes a consumer of new information to believe it is true if it aligns with their current beliefs or ideology\cite{kim2019combating}}, Believability \cite{french2023impact} and topic familiarity\cite{azzopardi2021cognitivebias}.We are uncertain how pre-existing beliefs affect the effectiveness of auditory warnings. This leads us to our next question: how do user's pre-existing beliefs/cognitive biases impact the effectiveness and perception of such auditory warnings?

In summary, we discussed the use of auditory interventions to address misinformation in podcasts. We proposed focusing on monologue-style podcasts and examining how auditory warnings impact user comprehension and the ability to detect misinformation. We then discussed the different dimensions (i.e., types and placement) of auditory warnings, including their effects on user comprehension, and the potential influence of pre-existing beliefs or cognitive biases. In light of this background knowledge, we put forth the following open challenges:

\begin{enumerate}
  \item Can auditory warnings included in podcasts assist users in recognizing misinformation?
  \item What kinds of auditory warnings are more effective in helping users be aware of misinformation? Do they impact the user’s comprehension of the information? Which ones are most salient and consistently perceived by different users? What is their impact on user engagement?
  \item How does the position of auditory warnings in a podcast w.r.t the snippet impact the user's comprehensibility and misinformation detection?
  \item Without any prior training, how do users perceive these auditory warnings? I.e., Can users conceptually map the auditory warnings to the presence of misinformation?
  \item Does users' pre-existing beliefs/cognitive biases impact the effectiveness and perception of such auditory warnings?
\end{enumerate}    
\section{Proposed Approach}
\label{sec:setup}

We aim to foster discussions among researchers and practitioners by outlining open research challenges and emphasizing future steps in exploring auditory interventions against misinformation in podcasts. However, conducting experiments to study their effectiveness requires controlling multiple variables (see Table \ref{tab:variable-descriptions}). Moreover, such studies require controlling for potential effects from podcast content characteristics, such as topic complexity, political or societal stance of speakers, language comprehensibility, and podcast length. In the next section, we propose controlling these variables by simulating a podcast experience using synthesized content and detailing the approach researchers and practitioners could adopt.

\subsection{Data and Instruments}
\label{subsec:data_instruments}
This section discusses the topics used in the experiments, the process used to generate podcast content, and the auditory icons used.

\begin{table*}[th]
\small
\centering
\caption{Experiment variables to be considered while studying the effectiveness of auditory warnings along with the rationale.}
\label{tab:variable-descriptions}
\vspace{-1em}
\adjustbox{max width=\textwidth}{
\begin{tabular}{p{4.5cm}p{12cm}}
\toprule
\textbf{Variable} & \textbf{Rationale} \\
\midrule
1. Auditory Icon Type & {Section \ref{subsec:data_instruments}} \\
2. Position of Auditory Icon & {Section \ref{subsec:related_works_auditory_intervention}} \\

3. Comprehensibility & {Auditory feedback affects website usability~\cite{noh2018audioeffectwebcontent}, but their impact on podcast listeners is unclear.} \\
4. Misinformation Recall & {The user can be asked to identify the snippet they felt contained misinformation.}\\

5. Podcast Topic & {\citet{gadiraju2018knowledgegain}; \citet{gwizdka2006can}; \citet{li2008faceted}} \\
6. User Engagement & {\citet{noh2018audioeffectwebcontent}} \\
7. Icon recall rate & {Testing the user's recall of the auditory icons.} \\
8. Task Load & {\citet{Fazal2021InvestigatingCW};\citet{sanderson2022listening}} \\
9. PESQ score & {The score is an Automatic measure of sound quality \citet{loizou2011speech};\citet{Gohsen2023Nudge}} \\

10. Gender and Age & {\citet{martinez2023nobody}} \\
11. Language Proficiency & {\citet{muda2023people}} \\
12. Education Level & {A user's education level has been shown to have a positive impact on the recognition of fake news \cite{zrnec2022users}.} \\
13. Preferred Information Consumption & {A user's preferred mode of information consumption has shown to impact user engagement~\cite{rzepka2022voice}.} \\
14. Declared impairments & {Considering individual variations in those with brain injuries or neurological conditions~\cite{sitbon2023perspectives}.} \\
15. Speculated Meaning & {Determining users' perception of auditory icons can reveal if they can map auditory warnings to misinformation.}\\ 
\bottomrule
\end{tabular}
}
\end{table*}

\noindent \textbf{Podcast Topics and Content.}
We pick two topics with different levels of topical difficulty and potential knowledge gain~\cite{gadiraju2018knowledgegain} -- \textit{Altitude Sickness} ($T_1$) and \textit{Carpenter Bees} ($T_2$). To control for interaction effects between characteristics of the podcast speaker and users' perception of misinformation, we synthesize the content of the podcast using ChatGPT for both topics using the prompt: \emph{``Tell me about [INFORMATION NEED] using words a five-year-old would understand in a short paragraph of fewer than 200 words''} (see Table~\ref{tab:snippet-quotes}).\footnote{We refer to the information needs for \textit{Altitude Sickness} and \textit{Carpenter Bees} as discussed by \citet{gadiraju2018knowledgegain}.}
\begin{table}
    \small
    \centering
    \caption{Sample of a podcast content as generated by ChatGPT  -- GPT-3.5 on 21 November 2023 -- along with SSML tags to simulate a monologue podcast.}
    \vspace{-1em}
    \adjustbox{width=0.998\linewidth}{
    \begin{tabular}{p{2cm}p{4.5cm}}
    \toprule
    \textbf{Topic} & \textbf{Content} \\
    \midrule
        Altitude Sickness ($T_1$) & \verb|<speak>|
  \verb|<p>|Imagine you're flying in an airplane [\ldots] Sometimes, your body gets a bit sick.\verb|</p>|
  \verb|<p>|[\ldots] you can maybe take some medicine.\verb|</p>|
\verb|</speak>| \\
\midrule
        Carpenter Bees ($T_2$) & \verb|<speak>|
          \verb|<p>|There are special bees called carpenter bees. [\ldots] dead wood or bamboo.\verb|</p>|
          \verb|<p>|It's a bit tricky to tell [\ldots] tummies can tell the difference.\verb|</p>|
          
        \verb|</speak>| \\
    \bottomrule
    \end{tabular}
    }
    \label{tab:snippet-quotes}
    \vspace{-1em}
\end{table}

Next, we propose manually altering the snippet content to simulate misinformation, followed by generating an audio snippet for each topic using text-to-speech tools. \footnote{To test feasibility, we used the \textit{Amazon Polly Neural Engine} with \textit{English (US)} language and \textit{Joanne (Female)} voice.} Previous studies show that content length and comprehensibility can affect a user's information consumption behavior. Therefore, we propose controlling these factors in future experiments. For instance, by controlling the readability scores - Flesch-Kincaid ($F$)~\cite{flesch1948new} and Gunning Fog ($G$)~\cite{gunning2004plain} , word count ($WC$), audio length in seconds ($L$) across topics used in the study $T_1(F=7.18,G=9.34,WC=156,L=47)$ and $T_2(F=6.97,G=9.30,WC=155,L=50)$. Note that there is no major difference between the scores of $T_1$ and $T_2$. Using synthesized podcasts also ensures that variables like sound effects (e.g., intro/outro music, advertisements) in real-world podcasts don't interfere with the experimental variables.
Furthermore, we propose the experiments (where the participant would be facing the screen throughout the experiment) to exclude warning labels or visual indicators in the podcast interface to simulate a real-world scenario where users may not be visually interacting with the podcast~\cite{tobin2022people}.

\noindent \textbf{Auditory Icons.} To instantiate an experiment, we propose to start by focusing on two sub-categories of auditory icons (i) iconic/nomic and  (ii) metaphorical). Given the nature of iconic/nomic (i.e., the icon is closely related to the target) we propose to create one icon for each topic. Consequently, $T_1$ would have an icon that sounds like \lq \textit{Whispers in the wind}\rq. $T_2$ would have an icon that sounds like \lq \textit{Swarm of buzzing bees with distorted voices}\rq. Meanwhile, for metaphorical type, we propose to use only one type of icon for both the topics \lq \textit{Alarm bells} \rq.When creating these auditory icons, we emphasize keeping the loudness under 60 dB\footnote{Audio loudness is measured in decibels (dB). Typically, levels from 0 dB to 60 dB do not cause hearing damage with repeated exposure.} to prevent hearing impairments and avoid the unpleasantness and urgency associated with higher decibel levels \cite{mckeown2007mapping, momtahan1992mapping}.
\section{Summary}
\label{sec:conclusion}

We propose a novel approach to tackle misinformation in podcasts by harnessing auditory signals to alert listeners about potential inaccuracies within snippets of the podcast. By integrating such auditory signals we aim to empower listeners with real-time awareness. We identify and synthesize several research questions entailing this context.
We discuss several research gaps that warrant empirical investigation into the use of auditory signals for this purpose, and possibly in the future, the potential to personalize the experience of identifying misinformation with custom auditory icons. We believe 
this can inform the next steps to address the problem of tackling misinformation in audio-based content such as podcasts.
\begin{acks}
 This research is partially supported by the \grantsponsor{ARC}{Australian Research Council}{https://www.arc.gov.au/} (\grantnum{ARC}{CE200100005}, \grantnum{ARC}{DE200100064}) and the TU Delft AI Initiative. 
   
\end{acks}
\newpage
\balance
\bibliographystyle{ACM-Reference-Format}
\bibliography{99-references}

\end{document}